\documentclass[traditabstract]{aa}
\usepackage{txfonts}
\usepackage{graphicx}
\usepackage{natbib}
\usepackage{amssymb}

\newcommand{\kms}{\,km\,s$^{-1}$}	  
\newcommand{\kmsMpc}{\,km\,s$^{-1}$\,Mpc$^{-1}$~}	  
\newcommand{\lya}{Ly$\alpha$}
\newcommand{\es}{erg\,s$^{-1}$~}
\newcommand{\esc}{\mathrm{erg\,s}^{-1}\,\mathrm{cm}^{-2}}
\newcommand{\esca}{erg\,s$^{-1}$\,cm$^{-2}$\,arcsec$^{-2}$~}

\newcommand{\escan}{erg\,s$^{-1}$\,cm$^{-2}$\,\AA$^{-1}$~}

\newcommand{\msol}{M$_\odot$~}

\newcommand{\br}{BR~1033$-$0327}
\newcommand{\q}{Q~2139$-$4324}
\newcommand{\sdss}{SDSS~J0939$+$0039}
\newcommand{\ja}{SDSS~J14472$+$0401}
\newcommand{\jb}{SDSS~J21474$-$0838}
\newcommand{\brb}{BR~2237$-$0607}
\def\kmsmpc{km s$^{-1}$ Mpc$^{-1}$}

\begin{document}
\title{Spectroscopy of extended \lya\ envelopes around $z=4.5$ quasars
\thanks{Based on
observations made with the  FORS2 multi-object spectrograph mounted on
the Antu  VLT   telescope  at   ESO-Paranal   Observatory   (programme 
081.A-0538B; PI: P. North)}}
\author{P. L.~North\inst{1} \and F.~Courbin\inst{1}  \and A.~Eigenbrod\inst{1}
\and D.~Chelouche\inst{2}}
\offprints{P. L. North, \email{pierre.north@epfl.ch}}
\institute{Laboratoire d'astrophysique, 
           Ecole Polytechnique F\'ed\'erale de Lausanne (EPFL),
           Observatoire de Sauverny,
           CH-1290 Versoix, Switzerland
\and
           Department of Physics, Faculty of Natural Sciences,
	   University of Haifa, Haifa 31905, Israel}
\date{Received 4 June 2010 / Accepted 2 May 2012}
\authorrunning{North et al.}
\titlerunning{Extended Ly$\alpha$ envelopes around $z=4.5$ quasars}
\abstract{What are the frequency, shape, kinematics, and luminosity of \lya\
envelopes surrounding radio-quiet quasars at high redshift, and is the
luminosity of these envelopes related to that of the quasar or not? As a
first step towards answering these questions,
we have searched for \lya\ envelopes around six radio-quiet quasars at
$z\sim 4.5$, using deep spectra taken with the FORS2 spectrograph attached to
the UT1 of the Very Large Telescope (VLT). Using the multi-slit mode allows us
to observe several point spread function stars simultaneously with the quasar,
and to remove
the point-like emission from the quasar, unveiling the faint underlying \lya\
envelope with unprecedented depth.
An envelope is detected around four of the six quasars, which suggests that these
envelopes are very frequent. Their diameter varies in the range
$26\lesssim d \lesssim 64$~kpc, their surface brightness in the range
$3\times 10^{-19}\lesssim \mu\lesssim 2\times 10^{-17}$~\esca, and their luminosity
in the range $10^{42}\lesssim L(Ly\alpha) \lesssim 10^{44}$~\es.
Their shape may be strongly asymmetric. The \lya\
emission line full width at half maximum (FWHM) is $900 < FWHM < 2200$~\kms
and its luminosity correlates with that of the broad line region (BLR) of the quasar,
with the notable exception of BR2237-0607, the brightest
object in our sample. The same holds for the relation between the
envelope \lya\ luminosity and the ionizing luminosity of the quasar. 
While the deep slit spectroscopy presented in this paper is very efficient at
detecting very faint \lya\ envelopes, narrow-band imaging is now needed to
measure accurately their spatial extent, radial luminosity profile, and total luminosity.
These observables are crucial to help us discriminate between the three possible
radiation processes responsible for the envelope emission: (i) cold accretion,
(ii) fluorescence induced by the quasar, and (iii) scattering of the BLR
photons by cool gas.}


\maketitle

\section{Introduction}

The so-called ``\lya\ blobs'' have attracted much attention in the past few
years. They are extended nebulae at high redshift ($z\sim 2-5$) emitting in the
\lya\ line, with typical sizes of $\sim 10\arcsec$ or $\sim 100$~kpc
\citep[]{SAS00, MYHT04, MYH11, YZTE09}. Their \lya\ luminosity is typically
$\sim 10^{43}$~\es, and, while they are numerous at high redshift, they seem to
disappear at moderate to low redshift \citep{KWCW09}. The source of their power
has remained unclear and a controversial subject. Some host an active galactic
nucleus (AGN) at their
centre, while others apparently do not \citep[]{GALS09, YZTE09}.
Recently, extended \lya\ emission was discovered around the $z=6.4$
radio-quiet quasar CFHQSJ2329-0301 \citep[]{WCB11, GUW12}.

Extended \lya\ emission has also been observed around high-redshift radio
galaxies \citep[]{H91a, vORC96}, as well as around radio-quiet quasars
(\citealt{CJW06}; \citealt{BSS03}). The \lya\ flux from the nebula
is about 0.5\% of the flux in the
integrated broad \lya\ line of the QSO, in the case of radio-quiet quasars (RQQs).
For radio-loud quasars (RLQs) or radio-galaxies, the \lya\ flux of the nebula is
an order of magnitude higher \citep{CJW06}, presumably because the  emission of
RLQ  gaseous envelopes is enhanced by interactions with the radio jets. In
addition, it seems that the RLQs generally reside in richer environments than
RQQs \citep[]{EYH02}.

Two main power sources have been suggested to explain the luminosity of the
\lya\
blobs (excluding the envelopes of RLQs). The first is photo-ionization by a
central source such as an AGN or a starburst region \citep[e.g.][]{CMBG08}.
\cite{GALS09} recently
presented arguments in favour of this idea, based on deep X-ray
observations of \lya\ blobs. They unveiled obscured AGNs but found no diffuse
X-ray emission that would have betrayed the existence of a hot gas component
with $T\gtrsim 10^7$~K; the lack of such a component rules out inverse Compton
scattering of CMB photons as an ionizing source. The other possible power source
is cold accretion, a
scenario where pristine gas falling into the potential well of a dark matter
halo gets heated and ionized by collisions, converting gravitational energy into
\lya\ radiation. According to \cite{DL09}, observed \lya\ blobs could be explained
in this way if only $\gtrsim 10$\% of the gravitational binding energy of cold
gas being accreted onto a galaxy is converted
into \lya\ radiation (such a conversion was also discussed by
\citealt{HSQ00} and by \citealp{FKG01}). In addition, since the emission would come from
filamentary cold flows, it would not depend on the existence of a central
ionizing source such as an AGN, because the cold gas would be self-shielded from
the ionizing radiation.

A blind search for \lya\ blobs has the advantage of providing a bias-free sample,
but is very time-consuming. It requires wide-field, narrow-band imaging, and the
candidates found in this way still have to be observed spectroscopically to be
confirmed. Alternatively, looking for \lya\ blobs around radio-quiet quasars
takes advantage of pre-existing catalogues. In addition, bright quasars are
thought to reside in massive halos ($\sim 5\times 10^{12}$~\msol, see e.g.
\citealp{HLH07}), which are also thought to host the so-called ``cold flows''
\citep{KKF09} that could emit a substantial \lya\ flux \citep{DL09}.
The main drawbacks of this alternative search technique are that (1) it
is not guaranteed {\sl a priori} that a \lya\ nebula is present around each
quasar, (2) the sample will inevitably be biased, and (3) the bright quasar emission
has to be subtracted before one can see and characterize the \lya\ envelope.

We initiated, a few years ago, a pilot survey aimed at exploring  in 
detail the spatial  extent, the luminosity  and kinematics of the large
hydrogen  envelopes of remote quasars spanning a broad magnitude range, at
redshift $z\sim 4.5$. In this way, the possible relation between the  quasar
luminosity and  the extent and luminosity of the envelope could be explored.
In a model where photoionization by the AGN is the main cause of the \lya\
luminosity \cite[e.g.][]{HR01}, one should expect the latter to be strongly
correlated with the luminosity of the quasar, at least that coming from the
ionizing Lyman continuum. Alternatively, in the model of \cite{DL09}, no direct
relation between the AGN luminosity and that of the blob is expected.
Preliminary results of our survey have been published \citep[Paper~I]{CNEC08}
for three
quasars, one of which is surrounded by a \lya\ envelope (it appears that the
marginal detection of a second one was spurious); the observational
strategy was explained in more details in that paper. The   present article
describes our
detection of a \lya\, nebula around two more quasars, and a marginal detection
around a sixth one. The observations are
described in Section \ref{observations} and the results are presented in
Section \ref{results}. The results are discussed assuming a cosmology with
$H_0=72$~\kmsMpc, $\Omega_M=0.3$, and $\Omega_\lambda =0.7$.

\begin{table}
\caption{Journal of observations and main characteristics
of  the  quasars.}
\label{journal}
\centering
\begin{tabular}{ccccc}
\hline\hline
 Name     &JD(start)&Exposure&Airmass& Seeing	   \\
	 &$-2400000$ &time (s)&(start)& (")	   \\ \hline
\multicolumn{5}{l}{SDSS\,J14472+0401, z=4.580, R(AB)=19.9$\pm$0.2, M$_R$=-28.2} \\
\hline 
      &54647.617& 1300   & 1.30 &$0.54$\\
      &54647.633& 1300   & 1.38 &$0.57$\\
      &54653.565& 1300   & 1.18 &$0.61$\\
      &54653.581& 1300   & 1.22 &$0.60$\\
      &54654.562& 1300   & 1.18 &$0.85$\\
      &54654.577& 1300   & 1.22 &$0.88$\\
      &54654.600& 1300   & 1.31 &$0.89$\\
      &54654.615& 1300   & 1.39 &$0.90$\\ \hline
\multicolumn{5}{l}{SDSS\,J21474-0838, z=4.588, R(AB)=18.7$\pm$0.2, M$_R$=-29.3} \\
\hline 
      &54645.875& 1300   & 1.07 &$1.04$\\
      &54645.891& 1300   & 1.10 &$0.99$\\
      &54647.700& 1300   & 1.48 &$0.93$\\
      &54647.715& 1300   & 1.35 &$0.90$\\
      &54647.736& 1300   & 1.23 &$0.69$\\
      &54647.752& 1300   & 1.17 &$0.91$\\
      &54647.772& 1300   & 1.11 &$0.55$\\
      &54647.787& 1300   & 1.08 &$0.57$\\ \hline
\multicolumn{5}{l}{BR\,2237-0607, z=4.550, R(AB)=18.3$\pm$0.2, M$_R$=-29.8} \\
\hline 
      &54653.750& 1300   & 1.29 &$0.69$\\
      &54653.766& 1300   & 1.21 &$0.68$\\
      &54653.790& 1300   & 1.13 &$1.02$\\
      &54654.716& 1300   & 1.52 &$0.97$\\
      &54654.732& 1300   & 1.39 &$1.14$\\
      &54654.751& 1300   & 1.27 &$1.26$\\
      &54654.767& 1300   & 1.20 &$1.18$\\
      &54656.803& 1300   & 1.09 &$0.82$\\
      &54656.818& 1300   & 1.07 &$0.91$\\ \hline
\end{tabular}
\tablefoot{The apparent   magnitudes are   given  in  the  AB
system. They are computed by integrating the quasar's spectrum through
the {\tt RSPECIAL} ESO filter  curve.  The absolute magnitude  assumes
$H_0=72$   \kmsmpc and $(\Omega_m ,\Omega_{\lambda})=(0.3,0.7)$.}
\end{table}

\begin{figure}
\resizebox{\hsize}{!}{\includegraphics{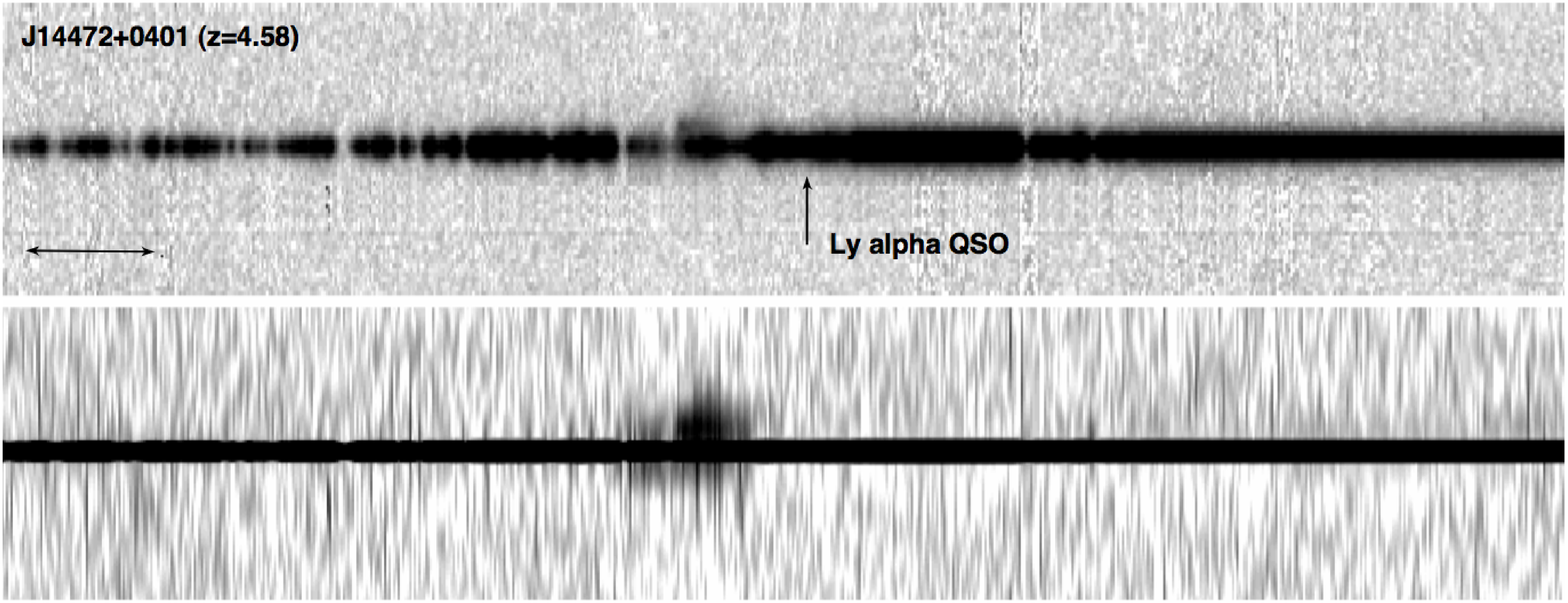}}\vskip 1pt
\resizebox{\hsize}{!}{\includegraphics{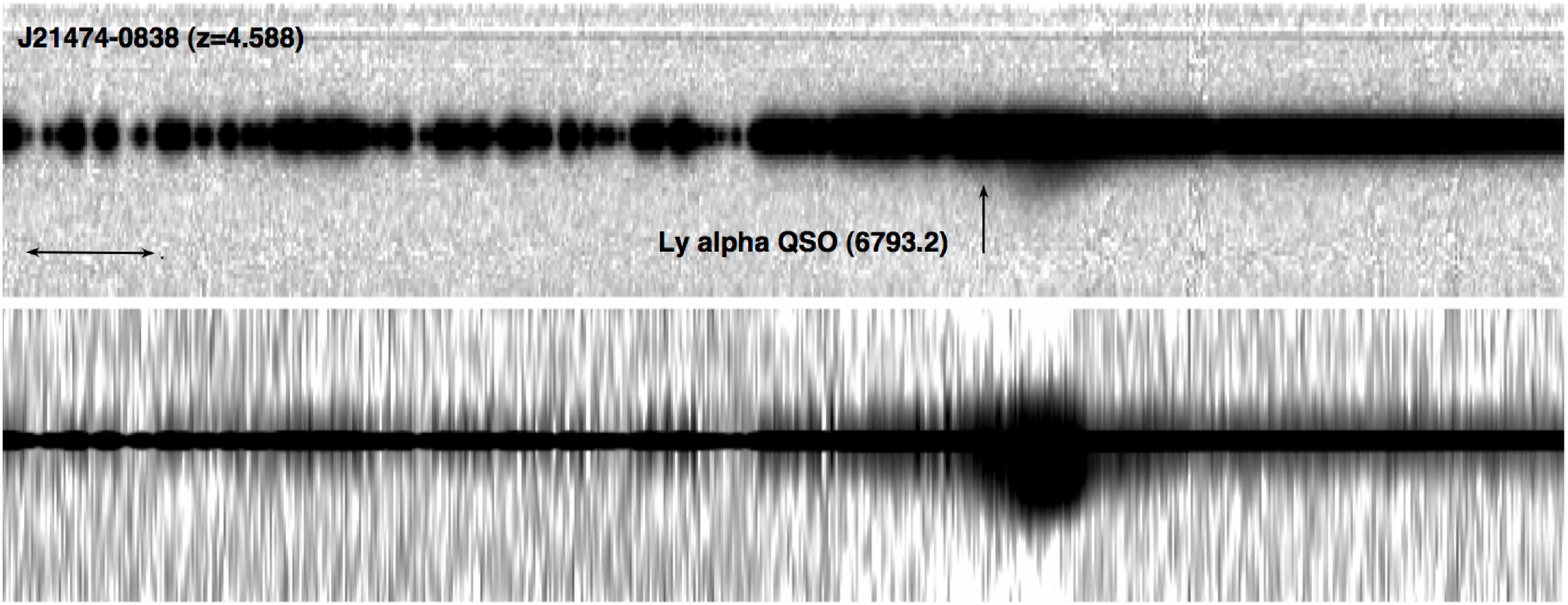}}\vskip 1pt
\resizebox{\hsize}{!}{\includegraphics{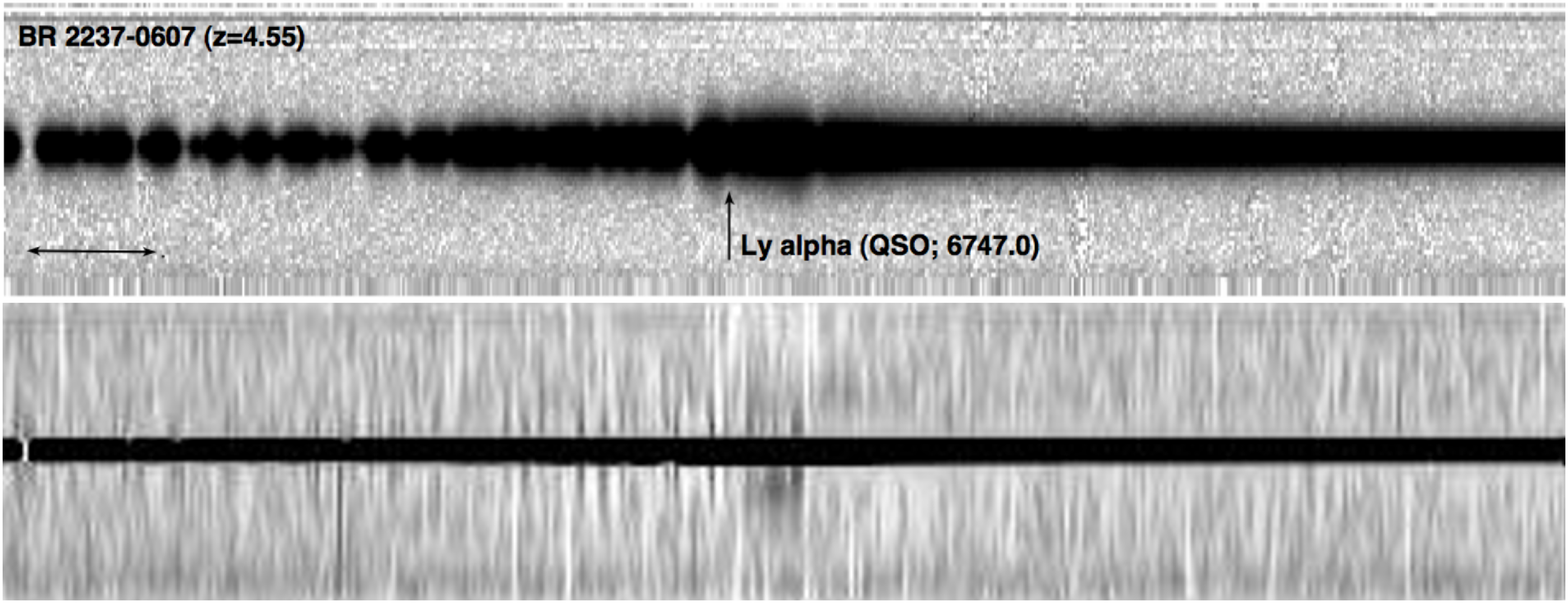}}
\caption{The three combined and sky-subtracted FORS2 spectra along
with their  spatial deconvolution.     The spatial  resolution   after
deconvolution  is  0.25\arcsec.  In   each  panel, the  vertical arrow
indicates  the  position of  the  \lya\  emission  line of the
quasar, the wavelength of which is indicated. The horizontal double arrow
is 50 \AA\ wide. A \lya\  envelope is detected  in all three
quasars. The envelope of \ja\ is blueshifted from that of the 
quasar, while that of \jb\ and that of \brb\ are redshifted. The height of
the spectra is 16\arcsec\ in all panels.}
\label{2Dspectra}
\end{figure}

\section{Deep VLT optical spectroscopy}
\label{observations}

\subsection{Sample and observations}
The quasars were selected from the 5th edition of the SDSS catalogue of
quasars \citep{SHRS07} and from the 12th edition of the V\'erons'
catalogue \citep{VV06}. All six objects are listed in Table\_QSO of
the V\'erons' catalogue with an asterix before their name, which means that
they have not been explicitly associated with a radio source in the literature.
One of them, BR1033-0327, was explicitly considered as radio-quiet by
\cite{KBTV08}. The other objects that were optically detected
are probably radio quiet, though statistically a possibility remains that
one of them might be radio-loud. The fraction of radio-loud objects
in quasars emitting close to the Eddington limit is indeed $\lesssim 5$\%
\citep[Fig.~9]{RCW09}; thus, there is at most one
chance in three that one of our six quasars is radio-loud.

The  observations of the three new targets were carried  out  in ESO Period 81
(April-September 2008) in service  mode  with the FORS2 multi-object
spectrograph
attached to  VLT-UT1.  The ESO  grism G1200R+93 has a  resolving power
$R=1070$ with a  2\arcsec-slit, which ensures that we can catch most of the flux of
the \lya\, envelope.   This  grism is  used  in  combination with  the
GG435+81 order separating  filter, leading to the wavelength  coverage
6000 \AA\  $< \lambda  <$  7200 \AA.   The maximum efficiency  of this
combination coincides well  with   the  expected wavelength of     the
redshifted \lya\ line, i.e. about 6686 \AA.

The multi-slit  MXU mode  is used,  with  slits that are   long enough
(typical  length: $\sim 20\arcsec$) to reliably  model and subtract the sky
emission.  Only one scientific target  is observable in each
field,  but the  MXU capability  is used  to observe several stars
through identical  slits. In this way, a spectral point spread function
(PSF) is measured
simultaneously with  the  quasar.  This  is  crucial  for the  spatial
deconvolution of the data  to work efficiently \citep{COU2000} and  to
clearly separate the quasar spectrum from that of the putative envelope.

To properly remove cosmic rays, the  total exposure time  is
split   into several shorter  exposures.  The   journal  of  the actual
observations is presented in Table~\ref{journal}.

\subsection{Reduction and spatial deconvolution}

The data  reduction was carried out using the
standard  {\tt  IRAF} procedures.  The   individual  spectra listed in
Table~\ref{journal} were flat-fielded using dome flats,   and
wavelength-calibrated in two dimensions  in  order to correct the  sky
emission lines for  slit curvature. The  scale of the reduced  data is
0.76  \AA\, per pixel in  the  spectral direction and 0.25\arcsec\, in
the spatial direction.

The   sky emission was then  subtracted from the  individual frames by
fitting a  second order polynomial along  the spatial direction.  This
fit considered only ten pixels on each side of  the slit, and the sky
at the position of the quasar was interpolated using this fit, both on
the quasar and the PSF stars.

The   cosmic rays were   removed using  the {\tt L.A.Cosmic}  algorithm
\citep{vanDokkum2001}. All the frames were visually checked after this
process to ensure that  no  signal was mistakenly removed
from the data.

The shape
of the spectra along  the spectral direction is slightly distorted,
i.e.,  the  position  of the   spectrum  changes   as  a function   of
wavelength.  These distortions  are corrected for  and the spectra are
eventually  weighted so  that their  flux is  the  same at a reference
wavelength, before they are combined to form a deep two-dimensional (2D)
spectrum. We show
in Fig.~\ref{2Dspectra}   the   combined  spectra  of  the  three new
quasars,    after   binning in both the spatial and  the spectral
directions.   An  extended \lya\, envelope  is  already  visible in
all three objects.

We spatially deconvolved  the  spectra
following the method described in \citet{COU2000}, which is an
adaptation to   spectroscopy  of  the  ``MCS''  image    deconvolution
algorithm \citep{MCS}. The results of  this deconvolution are
displayed  in  Fig.~\ref{2Dspectra}.  The  algorithm uses  the spatial
information contained in the spectrum of several PSF stars in order to
sharpen the data in the  spatial direction. At  the same time, it also
decomposes the  data  into   a point-source and     an extended-source
channel.  The output of  the deconvolution  procedure consists of  two
individual spectra, one for the quasar and one for its host galaxy (or
\lya\,   envelope),   free of any  mutual  light   contamination. It is
therefore  possible to estimate the luminosity  of  the \lya\, emission
``underneath''  the quasar. Subsampling of   the data is also possible
with the  MCS algorithm, hence  the pixel size in Fig.~\ref{2Dspectra}
is half that of  the   original data,  i.e.,  the  new pixel size   is
0.125\arcsec.

\begin{figure*}
\centering
\includegraphics[width=7cm]{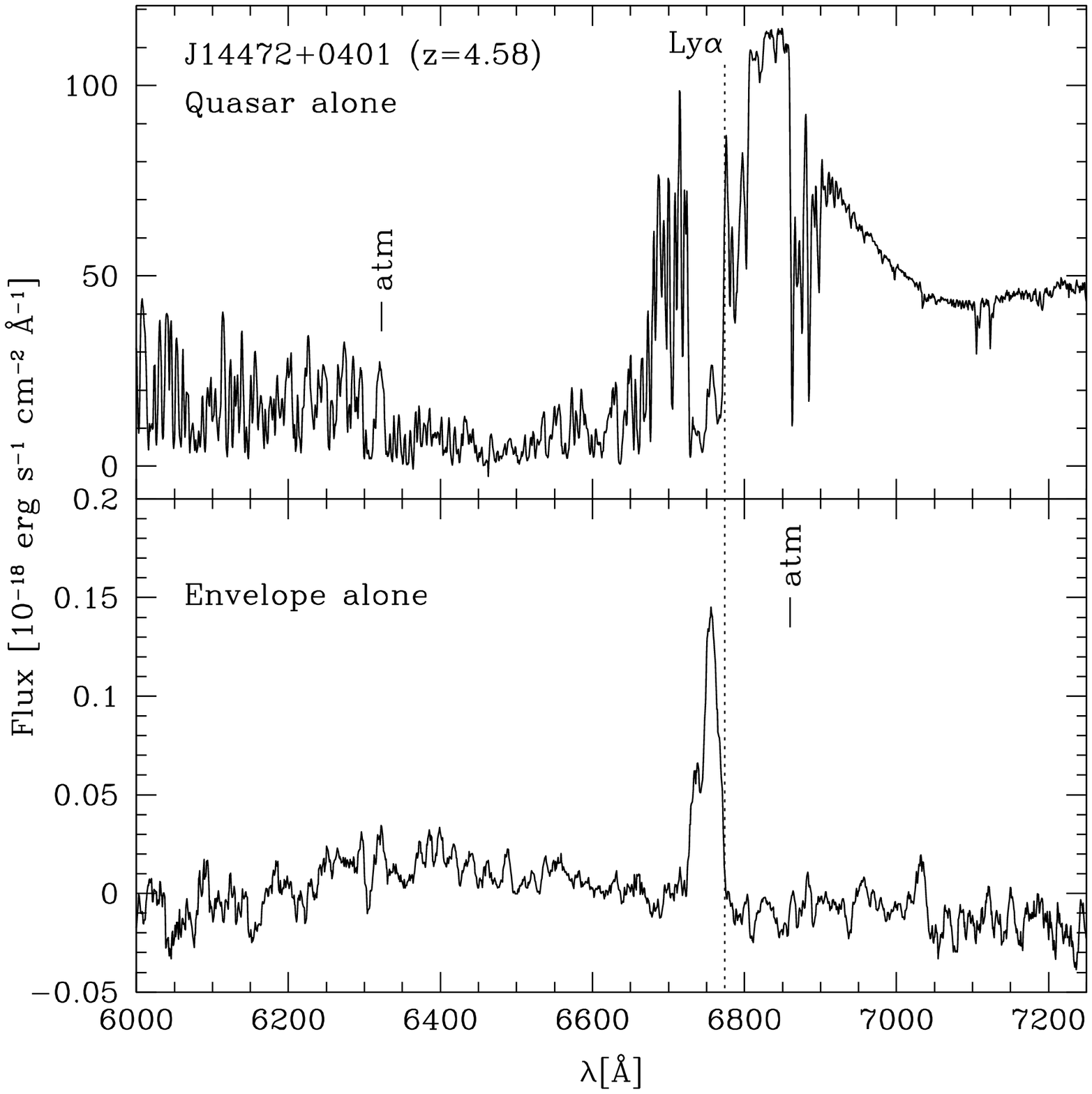}
\includegraphics[width=7cm]{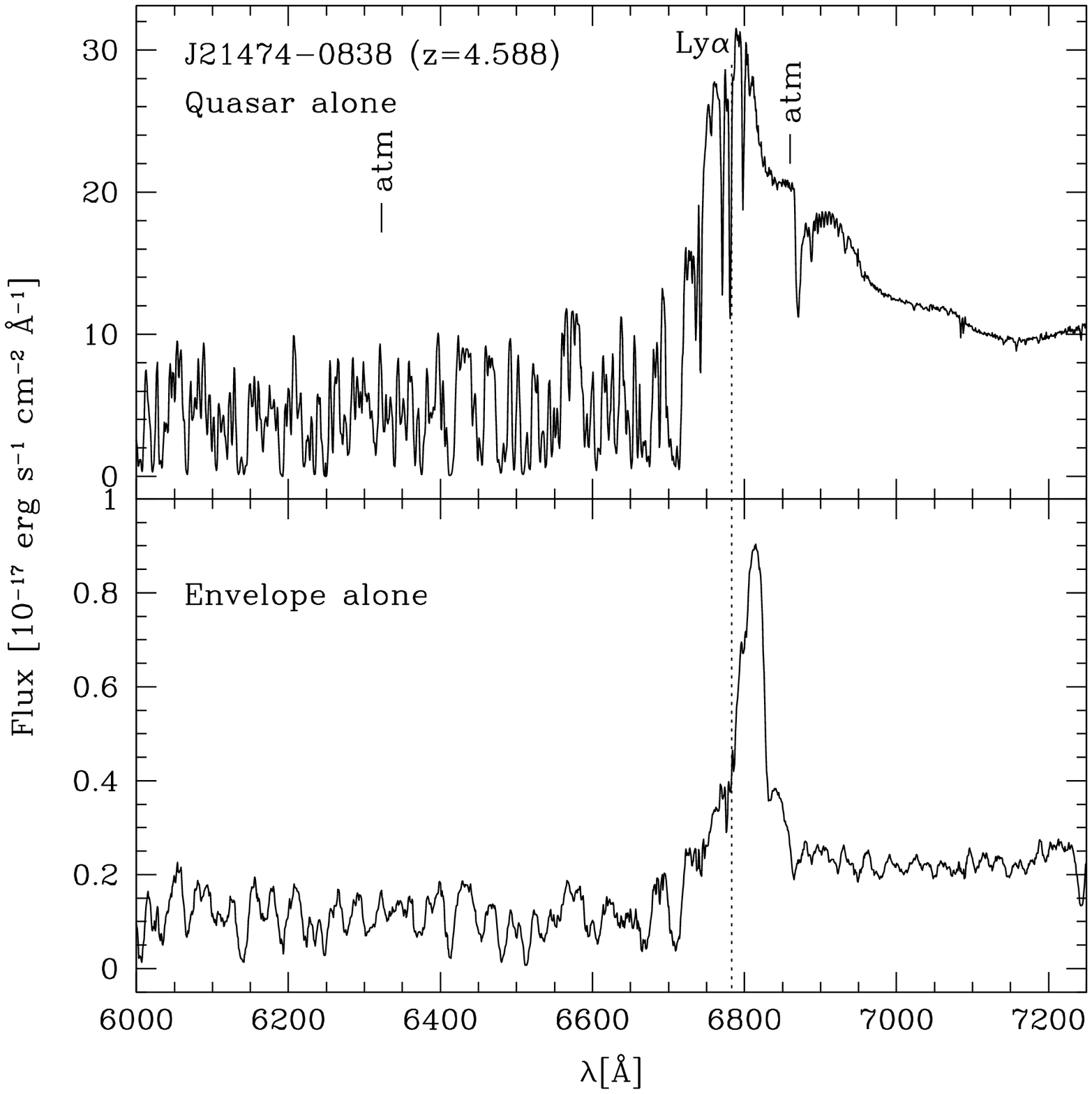}\\
\includegraphics[width=7cm]{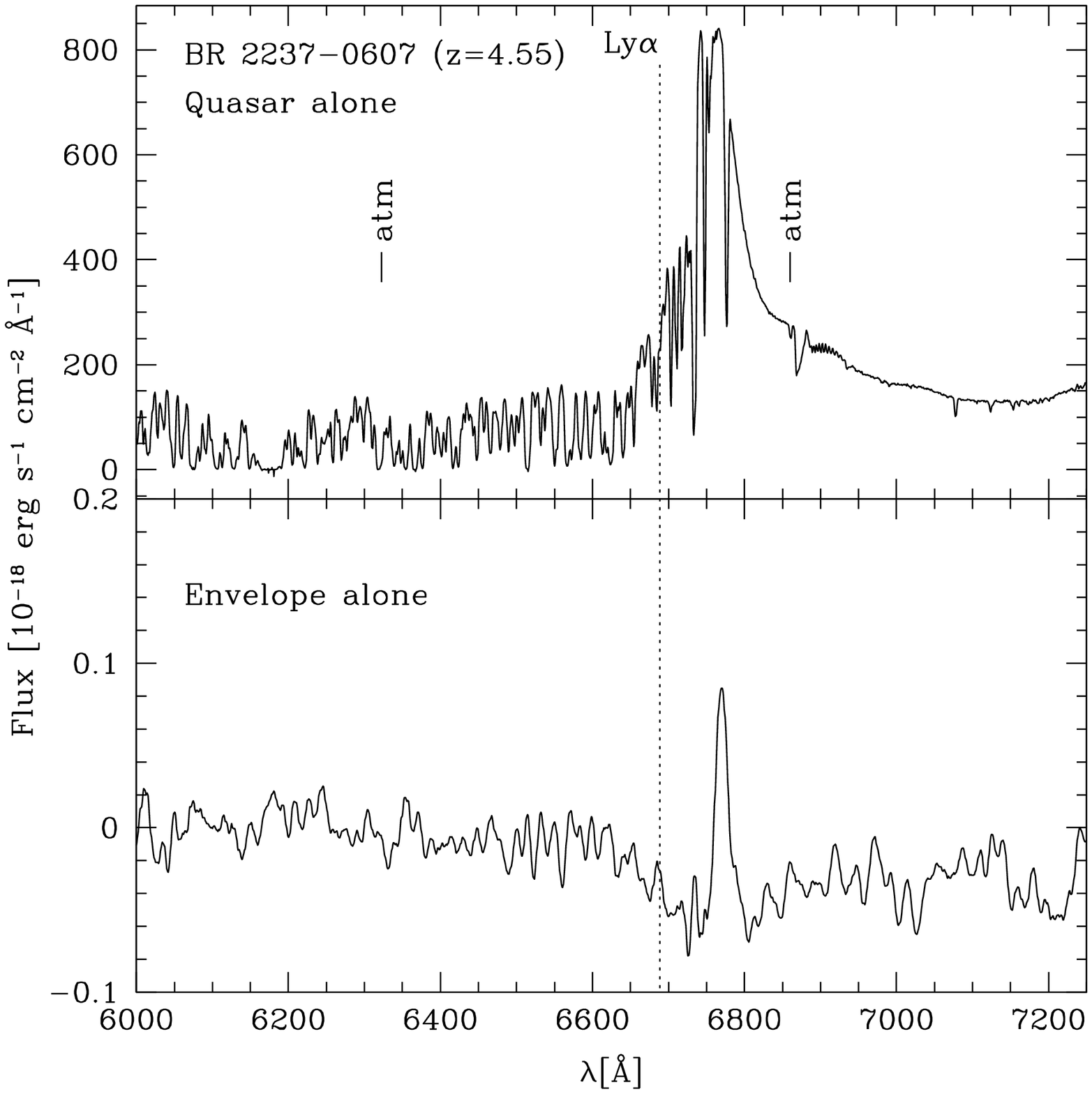}
 \caption{Extracted 1D spectra  for the three  quasars.  In each case,
 the top  panel shows the quasar alone  and the bottom panel shows the
 \lya\, envelope  alone, after  spatial  deconvolution of the  spectra.
 The vertical dotted  line  indicates the position of  the un-absorbed
 \lya\, emission line, at the redshift of the quasar (see subsections 3.1 and 2.1).  In
 all cases, the spectrum of the \lya\, envelope has been smoothed using
 a  boxcar of   8  \AA.} 
\label{1Dspectra}
\end{figure*}

\begin{table*}[t!]
\caption{Main properties of the \lya\, envelopes for all six quasars measured to
date (Paper~I and this work).}
\label{lyman}
\centering
\begin{tabular}{lcccccc}
\hline\hline
Object & $\lambda$     &     FWHM  &Extent&   Integrated flux            &  $3-\sigma$ detection limit   &   Surface brightness\\
       & [\AA] &[\AA]&(\arcsec; kpc)&[erg s$^{-1}$ cm$^{-2}]$& [erg s$^{-1}$ cm$^{-2}$]&[erg s$^{-1}$ cm$^{-2}$ \arcsec$^{-2}$]\\ 
\hline
\sdss&         $-$   &   $-$ &$(7;45)$&    $-$                      & $3.9\times10^{-18}$\tablefootmark{a} &$<2.8\times 10^{-19}$\tablefootmark{a}\\
\br &$6725.0 \pm 0.5$&$50\pm10$&10;64 &$1.4(\pm 0.1)\times 10^{-17}$& $6.7\times 10^{-18}$ &$7.2(\pm 0.5)\times 10^{-19}$\\
\ja &$6756.3 \pm 0.5$&$30\pm 3$& 6;38 &$4.3(\pm 0.7)\times 10^{-18}$& $2.5\times 10^{-18}$ &$3.6(\pm 0.6)\times 10^{-19}$\\
\q  &$6641.0 \pm 0.3$&  $-$   &$(7;45)$&  $-$                      & $2.2\times 10^{-18}$\tablefootmark{a} &$<1.6\times 10^{-19}$\tablefootmark{a}\\  
\jb &$6808.2 \pm 1.0$&$50\pm 8$& 8;51 &$3.1(\pm 0.2)\times 10^{-16}$& $3.3\times 10^{-18}$ &$1.9(\pm 0.1)\times 10^{-17}$\\  
\brb&$6773.6 \pm 2.0$&$19\pm 2$& 4;26 &$2.6(\pm 0.8)\times 10^{-18}$& $2.4\times 10^{-18}$ &$3.3(\pm 1.0)\times 10^{-19}$\\
\hline
\end{tabular}
\tablefoot{
All parameters  are given  in  the
observer's frame.  The $3-\sigma$ detection limit is based on the standard
deviation of the background noise according to equ.~\ref{LemLBLR}. The
surface  brightness is integrated in wavelength but given per arcsec$^2$,
while   the $3-\sigma$ limit is spatially and spectrally integrated.\\
\tablefoottext{a}{assuming an average extent of 7\arcsec}
}
\end{table*}

\begin{table*}[t!]
\caption{\lya\, luminosity of the quasars in the BLR, 
compared with the luminosity of  the extended envelopes.}
\label{lyman_lum}
\centering
\begin{tabular}{lccc}
\hline\hline
Object   &     L(BLR)     &      L(\lya)    &L$_\mathrm{corrected}$(\lya) \\
         & [erg s$^{-1}$] &  [erg s$^{-1}$] &  [erg s$^{-1}$] 	\\
\hline
\sdss&$4.1(\pm 0.2)\times 10^{44}$&$<7.4\times 10^{41}$\tablefootmark{a}&$<2.0\times 10^{42}$\tablefootmark{a} \\
\br  &$7.2(\pm 0.4)\times 10^{45}$&$2.7(\pm 0.2)\times 10^{42}$&$(1.1\pm 0.1)\times 10^{43}$ \\
\ja  &$2.1(\pm 0.1)\times 10^{45}$&$8.6(\pm 1.4)\times 10^{41}$&$(2.0\pm 0.3)\times 10^{42}$ \\
\q   &$4.6(\pm 0.2)\times 10^{44}$&$<4.1\times 10^{41}$\tablefootmark{a}&$<1.1\times 10^{42}$\tablefootmark{a} \\
\jb  &$6.8(\pm 0.3)\times 10^{45}$&$6.2(\pm 0.4)\times 10^{43}$&$1.9(\pm 0.1)\times 10^{44}$ \\
\brb &$1.4(\pm 0.1)\times 10^{46}$&$5.1(\pm 1.6)\times 10^{41}$&$8.0(\pm 2.5)\times 10^{41}$ \\
\hline
\end{tabular}
\tablefoot{
The flux in the quasar BLR is measured in the wavelength interval 1200-1230\AA\
(rest-frame). The last column gives the \lya\,  flux of the envelope,
after correction for slit-clipping (see text). The objects published in
\citet{CNEC08} are included for the sake of completeness.\\
\tablefoottext{a}{assuming an average extent of 7\arcsec}
}
\end{table*}

\section{Results}
\label{results}
The spectra of the three newly found \lya\ envelopes are shown in Fig.
\ref{1Dspectra} together with the spectra of the corresponding quasars
(see Fig. 4 of \cite{CNEC08} for the three other quasars previously observed).
The \lya\ luminosity
of the   envelopes,  their  angular size,   and   their  mean  surface
brightness are  presented     in  Tables~\ref{lyman}  and
\ref{lyman_lum}. We integrated the deconvolved
spectra  in the FORS2 {\tt  RSPECIAL} filter, which
is  also used to obtain short  acquisition images prior  to the
long  spectroscopic exposures.   These ``spectroscopic'' AB magnitudes
are  given in  Table~\ref{journal}.  We checked that they are compatible
with  the simple aperture  photometry obtained from  the short  images.

A \lya\ envelope was detected in four objects out of six, which suggests
that these envelopes are present in two thirds of the QSOs at $z\sim 4.5$.
Our  flux limit (integrated over the whole slit and \lya\ spectral profile)
is indicated in Table~\ref{lyman} and discussed below.
The lack of envelope around \sdss\ and \q\ may be real, but could also
be due to an unfortunate orientation
of the slit, if the latter is oriented perpendicular to the long axis
of the envelope.

\begin{figure}
\resizebox{\hsize}{!}{\includegraphics{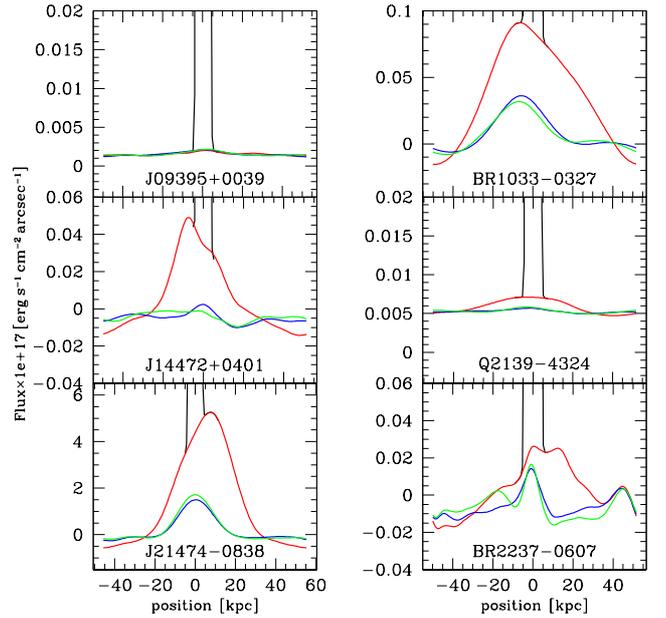}}
\caption{Spatial profiles of the envelopes. The black, quasi-vertical lines show
the intensity profile of the quasar, while the red line is the intensity profile
of the envelope, integrated across the spectral width of the \lya\ line (i.e. an
arbitrary range of $20$~\AA\ for objects with no significant envelope,
and a $1\times FWHM$ for the other
objects). The blue and green lines show the noise, i.e. the profile integrated
across wavelength ranges of the same width, but centred $76$~\AA\ bluer, respectively
redder than the envelope emission line (except for \brb, for which
the noise was sampled at $\pm 29$~\AA\ from the emission line).}
\label{profile}
\end{figure}

The measurable extent of the envelopes varies  (in terms of radius)
from $r\sim  20$~kpc for \brb, to
$r\sim  32$~kpc for  \br, measured from  the  quasar's centroid  to the
noise level in the spatial profiles shown in Fig.~\ref{profile}.

The surface  brightness of the  \lya\,  fuzz is unaffected  by
slit losses, but the total  luminosity   is.  Assuming that  the   \lya\,
envelopes are   uniform  face-on disks  with  diameters   equal to the
extents quoted  in Table~\ref{lyman},  we  can estimate  the amount of
flux  missed  by using   a   slit width  of 2\arcsec.    The  observed
luminosities of  the \lya\,  envelopes for  all four quasars are given in
Table~\ref{lyman_lum}, as well as the luminosities after correction
for the slit clipping.

\subsection{Detection limit}
\label{surf_bright_lim}
We now discuss the detection limit. This discussion
was not included in paper~I, where the detection limit was too optimistic
because it was based on the deconvolved images rather than on the original
ones. We now redefine this limit for all six objects observed so far
(three in Paper I, three in the present work, see Tables \ref{lyman} and
\ref{lyman_lum}).
We use the case of \brb\ for this discussion, because it is the target
with both the longest exposure time and the faintest envelope if any.

Using spectrophotometric standards, we verified
the reliability of the flux calibration by computing the
$R_\mathrm{special}$ magnitude of the quasar
from its observed spectrum, using the $R_\mathrm{special}$ filter transmission
curve: the obtained magnitude proved to be consistent with that obtained from the
imaging observations using the same filter. The maximum intensity of the quasar \brb\ is about
$370\,000$~ADU (analogic digital units), for one pixel along the wavelength axis
(representing $0.76$~\AA),
after summation along the spatial axis. Because the corresponding physical
flux is $8.41\times 10^{-16}$~\escan, this extracted pixel (i.e. the pixel of
the extracted quasar spectrum) receives a flux of
$6.39\times 10^{-16}$~erg\,s$^{-1}$\,cm$^{-2}$ on its $0.76$~\AA\ width.
Conversely, 1 ADU corresponds to
$1.73\times 10^{-21}$~erg\,s$^{-1}$\,cm$^{-2}$ in one pixel. We now consider an
extended source rather than a point source: the 8.23 pixels that cover the slit
width (considering only one pixel in the spatial direction) correspond to a
solid angle of $0.252\arcsec\times 2\arcsec=0.504\arcsec^2$ on the sky.
Therefore, there are $8.23/0.504=16.33$ pixels within a square arcsecond. Taking
into account an observed standard deviation of $\sigma_i\sim 40$ ADU (for individual
pixels) in the final 2D spectrum (corresponding to a total exposure time of
$11\,700$~s), one sees that the standard deviation in the total background
intensity, summed over a square arcsecond is (outside the sky emission lines)
\begin{equation}
\sigma(1\arcsec^2)\approx \sqrt{16.3}\times \sigma_i\sim 484~\mathrm{ADU}
\sim 8.4\times 10^{-19}~\esc ~\mathrm{.}
\end{equation}
This is also the error in a signal spread over 1 arcsec$^2$, provided that it
is small compared to the sky background (which is the case here) and assuming
that the latter is sufficiently well-defined to have a negligible error.
Therefore, the 3-$\sigma$ minimum signal that can be considered significant
after summation across the whole extent of the nebula (spanning $n$~arcsec$^2$) is
\begin{equation}
S(3\sigma)\simeq 12.1\times\sqrt{n}\times\sigma_i=484\,\sqrt{n}~\mathrm{ADU}
~\mathrm{.}
\label{LemLBLR}
\end{equation}

As a second consistency check, we used the Exposure Time Calculator for FORS2
(ESO version 3.2.7) and were able to
confirm the above figures: the flux
corresponding to 1 ADU is recovered to within 30\% (so the validity of our flux
calibration appears to be robust), and the standard deviation in the synthetic
2D spectrum, resulting from a single 11700~s exposure, is only slightly below
that of the empirical one ($\sim 30$~ADU instead of $\sim 35-40$~ADU).

In view of the above estimates, it appears that in paper~I, we were
too optimistic about the detection limit
in the case of \q: our present calculations imply that the envelope
cannot be assumed to have been detected, and
only an upper limit can be assigned to its flux.

\subsection{Comments on individual objects}
We now comment on briefly each of the three newly discovered objects.
\subsubsection{\ja} The extended emission is blue-shifted with
respect to the expected position of the \lya\ line of the quasar by
$\Delta \lambda = -20\pm 4$~\AA, i.e., $\Delta V=-889\pm 180$~\kms. The
redshift of the  quasar may be underestimated, as  it is only measured from
both the \lya\  line itself and the pair of \ion{C}{IV}~$\lambda 1549$ and
\ion{C}{III]}~$\lambda 1909$ lines, all of which are known  to  be
blueshifted with respect to the  actual redshift of  the quasar
\citep{Mintosh99}. The \lya\  line, on the other hand, is strongly affected by the
\lya\ forest. If the redshift of the quasar were underestimated, one would expect
the surrounding \lya\ emission to be redshifted with respect to the expected
position of the \lya\ line, while we see the reverse. In view of the strong
spatial asymmetry of the emitting region (see Fig. \ref{2Dspectra} and
\ref{profile}), it is
possible that the ``envelope'' is not centred at all on the quasar, but
is a blob at some distance away from it. The blue shift may be produced by
radiative transfer effects in both a collapsing neutral gas cloud
\citep{DHS06a,DHS06b} and shocks \citep{NM88}.

Interestingly, the nebular emission is matched by a deep absorption feature in
the quasar spectrum, at exactly the same wavelength. This is reminiscent of the
proximate damped \lya\ system (PDLA) discovered by \cite{HPKZ09} and associated
with the quasar SDSS J124020.91+145535.6, except that the \lya\ absorption does
not seem to be damped in our case. Some non-negligible flux remains at the
bottom of the line, and there is no indication of Lorentzian wings in the
absorption profile. Nevertheless, the width of this absorption is large, with
$FWHM=2100\pm 100$~\kms, which is about 50\% wider than the emission line of the blob;
the lack of damped wings suggests that the FWHM corresponds to the velocity
dispersion of the absorbing gas. On the other hand,
the emission line seems to have a double structure, which, if taken into
account, reaches about the same width as that of the absorption line. Higher
dispersion and higher signal-to-noise ratio spectra of this object would
allow us to clarify the latter point.

\subsubsection{\jb} This is by far the quasar with the brightest \lya\ blob in
our sample. It is an order of magnitude brighter than the second brightest one,
which surrounds the quasar \br. As in the latter object, the emission is slightly
redward of the expected position of the \lya\ line of the quasar, by
$\Delta \lambda = +32\pm 4$~\AA\ or $\Delta V=+1400\pm 180$~\kms. This is very
close to, though slightly larger than, the shift measured in \br, and the
explanation of it might be the same, i.e. the known blueshift of the \lya,
\ion{C}{IV}~$\lambda 1549$, and \ion{C}{III]}~$\lambda 1909$ lines.
Interestingly, the emission line appears to be slightly asymmetric, with a sharper
drop on its red side than the rise on its blue side; a similar asymmetry is seen
in the emission line of \br. This asymmetry might also be present in \ja, though it is less
prominent and possibly due to an absorption feature on the blue side of the
emission line.

The positive flux seen on Fig.~\ref{1Dspectra} on both sides of the envelope
emission line (and especially on its red side), is probably spurious. There is a
bad column on the CCD, running parallel to the quasar spectrum, about 28 pixels
away from it (or $7\arcsec$), which may have caused problems with the sky
subtraction. The quasar is so bright that a small imperfection in the PSF
profile might have produced a pseudo-continuum. 

\subsubsection{\brb} This quasar shows the faintest detected \lya\ blob in our
sample (disregarding the two non-detections), with a luminosity about two
times lower than the second faintest blob in our sample (\ja). This is also the
most difficult case, since the quasar is the brightest in the sample, so that
the contrast between the quasar and the envelope is the largest.
This is why it is detected only at the $\sim 3-\sigma$ level, according to the
criterion presented in Section 3.1. The emission is
strongly redshifted relative to the expected position of the \lya\ line of the
quasar: $\Delta \lambda = +87\pm 5$~\AA\ or $\Delta V=+3850\pm 220$~\kms. Such a
considerable shift cannot be explained by absorption, since the \lya\ line of the
quasar does show significant flux down to at least $6730$~\AA\, while the nebula
emits only between $\sim 6760$~\AA\ and $\sim 6790$~\AA. Thus, one has to admit that
either the redshift has been underestimated, or the nebula is receding at high
velocity from the quasar; it is unlikely that radiative transfer effects could
mimick such a fast motion, since one would require a high absorption/emission
cross-section in the line profile at these velocities.

\subsection{Redshift dependence}

The  observed surface brightness  of  the four \lya\, envelopes detected
in our programme is, on average, fainter by about 1-2 orders of magnitude
than for the CJW objects. Our brightest envelope has about the same 
surface brightness and integrated flux as typical CJW envelopes,
while our faintest ones are a hundred times dimmer. This contrast
cannot be ascribed exclusively to
the different redshift ranges  of the two  samples. Our sample is
at    $z_1 \sim 4.5$,  while most of the  objects   in  CJW are at
$z_2\sim 3.3$, so the flux ratio by redshift dimming alone would
be   $F(z_2)/F(z_1) = (1+   z_2)^4   / (1+z_1)^4\sim 2.7$,  i.e., much
smaller than the observed ratio.

Therefore, taken at face value, our sample as a whole suggests that higher-$z$
objects  tend to have lower  surface brightnesses. The total luminosities of
the envelopes in our sample overlap those in CJW: \jb\ has a higher \lya\
luminosity (when corrected for slit clipping) than all objects in CJW, the
envelope of \br\ has a luminosity close to the average one in the CJW sample, and
the two other objects are fainter. Thus, the envelope luminosities in our
sample seem to more closely match those of CJW than the surface brightnesses do, but
this may be due to our deeper sensitivity. Observations of the CJW envelopes
with the same sensitivity as ours would probably reveal that they have faint
extensions. Consequently, the envelopes would be found to have higher total
estimated luminosities, on the one hand, and a lower average surface brightness,
on the other. The spatial extent of the
envelopes in our sample ranges from $\sim 26$~kpc to $\sim 64$~kpc, while that of
CJW ranges from $\sim 10$~kpc to $\sim 60$~kpc. That explains why the luminosities
of our envelopes are often similar to those of CJW, in spite of a much lower
surface brightness. The  mean size of the envelopes in our sample is indeed
$\bar{r}\sim  45$\,kpc  compared  to $\bar{r}\sim  26.4$\,kpc in
CJW's sample. This translates into a ratio  of 2.9 in area.

The quasar luminosities are slightly larger in CJW's sample than in ours. The
average absolute magnitude $M_\mathrm{B}$, as listed in \cite{VV06}, is $-29.0$
for CJW's complete sample of seven quasars, while it is $-28.2$ for our sample of six
quasars. Since a correlation seems to exist between the luminosity of the quasar
and that of the \lya\ envelope (see below), such a magnitude difference may
explain part of the difference in \lya\ envelope luminosities.

A sample of quasars should be observed at $z\sim 3.3$ with the same
technique and depth as used at $z\sim 4.5$ in order to settle the question.

\begin{figure}
\resizebox{\hsize}{!}{\includegraphics{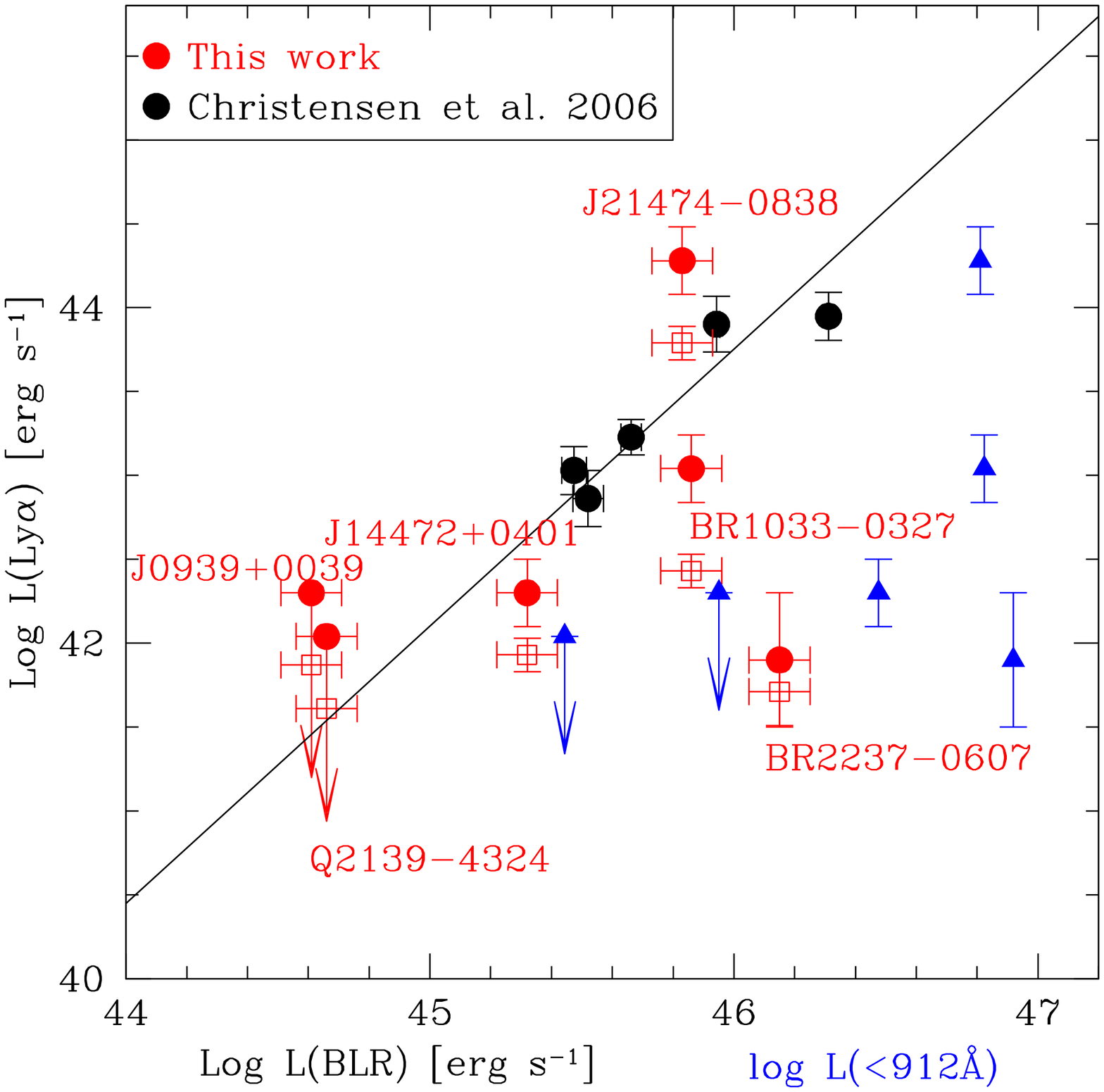}}
\resizebox{\hsize}{!}{\includegraphics{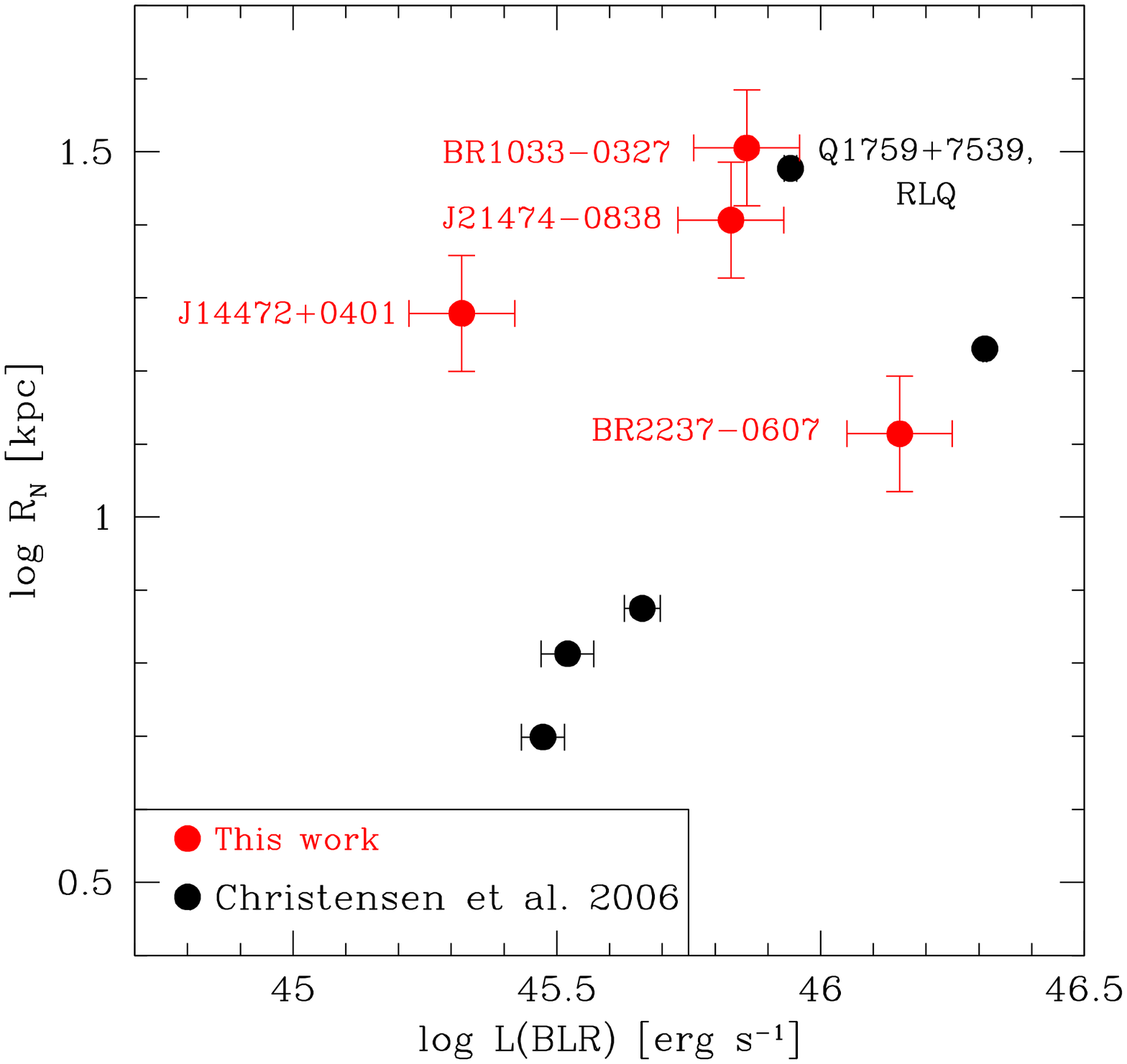}}
\caption{{\it Top:}  Relation between  the total luminosity  of 
the \lya\ envelope and that of the  quasar in the broad \lya\ line.
Our measurements  (red squares) are compared with CJW  (black
circles).  The open squares  represent the direct  measurements, while
the filled dots  are those corrected for slit clipping (see text).  The
downward arrows show our upper limits for
\sdss and \q, where no \lya\, envelope is seen.  In computing this limit, we assumed  
a radius r=23 kpc (i.e., the mean size of the other four objects) and we
did  not correct  for slit-clipping.
The black solid line was least squares fitted to the objects of CJW and to our
detected envelopes (with correction for slit clipping), except for \brb.
The blue triangles relate the \lya\
luminosity (corrected for slit clipping) with the luminosity of the ionizing
radiation from the UV continuum of the quasar ($\lambda < 912$~\AA, see
Subsection 3.3). The arrows again show the upper limits to the luminosity
of the undetected envelopes.
{\it  Bottom}: The radius of the
envelope as a function of $L(BLR)$.  Note the  tight trend followed by
the envelopes of the bright RQQs of CJW. Note also that  \br\, stands
above the maximum size of the RQQs of CJW, likely owing to our deeper
flux limit.}
\label{sb}
\end{figure}

\subsection{Dependence on the luminosity of the quasar}
The total flux in the \lya\, envelope is shown in Fig.~\ref{sb} as  a function
of the  quasar luminosity in  the broad \lya\, line. After correcting for
slit-clipping, we find that most points, combined with the ones of 
CJW, follow the linear relation
\begin{equation}
\log\left[\frac{L(Ly\alpha)}{L({\mathrm BLR})}\right]
=(0.65\pm0.53)\times\log[L({\mathrm BLR})]-(32.3\pm24.5)\,\mathrm{,}
\label{flux_env}
\end{equation}
suggesting that $L(Ly\alpha)$ varies strongly with $L({\mathrm BLR})$.
The regression line was determined after excluding
\brb\ (the two objects with upper limits to $L(Ly\alpha)$ were not
considered). The scatter is $0.39$~dex and the correlation coefficient is
$0.874$, leaving no doubt about the statistical significance. Nevertheless,
\brb\
remains two orders of magnitude below the linear relation, which seems difficult
to explain by slit-clipping effects alone. Since \sdss\ and \q\ may also lie well
below the regression line, the latter should perhaps be considered as an upper
envelope rather than as a real one-to-one relation proper.
However, the larger scatter of our points compared to those of CJW
around the regression line, may simply reflect our larger uncertainty in
L(\lya) due to the slit clipping and unknown shape of the envelope.
CJW did not have this problem because they used integral field spectroscopy.

The linear fit implicitly assumes no strong  redshift evolution of the
luminosity of the envelopes, since it relies on a mix of objects at redshifts
$z\sim 3$ and $z\sim 4.5$. Our new  observations  for the three
objects therefore seem to support the trend  that brighter quasars also
have  brighter  \lya\, envelopes,  under  the assumption of negligible
redshift evolution.

The  relation between   the   size of the   envelope  and  the  quasar
luminosity is  much less clear.  CJW find that brighter envelopes tend to be
larger, which, combined with the above correlation between envelope and BLR
luminosities, implies that brighter quasars should be embedded in larger
envelopes. We find no such trend on the basis of our results for our four objects,
as  shown  in  Fig.~\ref{sb} (bottom panel). The marginal trend obtained with the CJW data may
be explained by the difficulty in subtracting the quasar light: the brighter the
quasar, the wider the envelope has to be in order to be detected unambiguously.
Our deeper observations, coupled with a cleaner subtraction of the quasar light,
make our results less sensitive to this effect.
In addition, the small field of view  used in CJW  implies that there has been
some severe clipping  of  the   envelopes,  if they   extend  much   beyond  a few
arcsecs.

We now examine the possible correlation between the ionizing flux of the quasar
and the \lya\ flux of the envelope. As in \cite{CJW06}, we use the template
spectrum based on the Sloan Digital Sky Survey \citep{VBRB01}, completed by the
composite Far Ultraviolet Spectroscopic Explorer spectrum of \cite{SKB04}.
For simplicity, we adopted the power-law fit $F_\nu\propto \nu^{-0.56}$
determined by \cite{SKB04} to represent the
quasar spectrum in the far ultra-violet, from $800$~\AA\ to $912$~\AA. We redshifted
this composite template spectrum to the observed value of each of our observed
targets, and adjusted its intensity so that it matched the observed flux in the
range $\sim 1286-1291$~\AA\ in the rest frame ($\sim 7071-7099$~\AA\ at
$z=4.5$). Then, we integrated the spectrum between $800$~\AA\ and $912$~\AA\ after
having rebinned it to the rest frame, obtaining the ionizing luminosity
$L(<912$\AA$)$. This luminosity depends of course on the lower integration limit,
which is arbitrary. Another choice would have simply changed $L(<912$\AA$)$ by a
constant factor. Figure \ref{sb} (upper panel) shows the \lya\ luminosity of the
envelope (corrected for slit clipping) as a function of the ionizing
luminosity defined above. At first sight, there is no correlation, which tends
to confirm the result obtained by \cite{CJW06}, but at larger redshifts
($z\sim 4.5$) and for a wider range of \lya\ luminosities: about 2.3~dex
instead of $\sim 1$~dex, taking only the RQQs into account. However, the lack
of correlation is essentially due to \brb, and removing this object would result
in a clear trend, similar to the one found using the BLR luminosity instead of
$L(<912$\AA$)$. Hence, the interpretation of our results depends on the weight given to
\brb: if one fully takes it into account, then there is no correlation and
this could be interpreted as an argument against the quasar being the main cause
of the \lya\ emission of the envelope; conversely, excluding \brb\ as a
pathological case restores the correlation and leads to the opposite conclusion.
In any case, small number statistics and slit-clipping effects make such a conclusion
fragile, as well as the uncertainty inherent to the use of a uniform template
spectrum for the quasar.

\subsection{Surface brightness and width of the emission line}
The FWHM of the \lya\ emission line varies by only a factor of $\sim 2.5$ from
one object to the other in our sample, but it may be interesting to consider
whether it is correlated with e.g. the \lya\ luminosity of the nebula. The latter
spans a wide range, of about $2.3$~dex. Plotting $\log(FWHM)$ against
$\log L$(\lya), we indeed found a rough trend of increasing FWHM (corrected for the
intrumental width) with increasing luminosity, although there are four points only,
so it is not statistically significant. The correlation coefficient of 0.76 and
Student $t$ test of 2.03 leave about a 20\% 
probability of finding the observed correlation by chance alone.
Furthermore, the CJW data do not fit the correlation, which confirms that it
cannot be real.

\begin{figure}
\centering
\resizebox{\hsize}{!}{\includegraphics{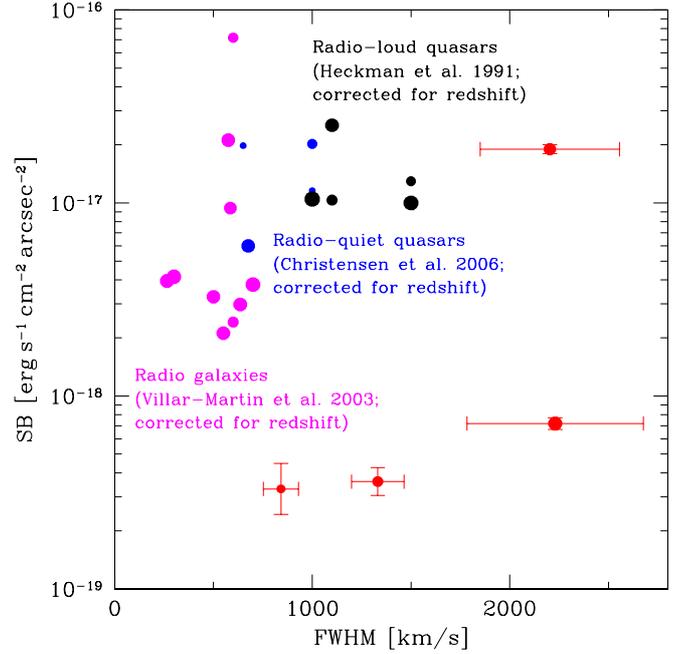}}
\caption{Average surface brightness of the envelope, as a function of the width
of the \lya\ emission line (expressed in \kms). The red dots represent our
results.
The surface of the points is proportional to the size of the envelope.}
\label{logSB_FWJM}
\end{figure}

Figure~\ref{logSB_FWJM} presents the average surface brightness of the envelope
versus the average FWHM of its \lya\ line, for our four objects as well as for the
RQQs studied by CJW, for the quiescent halos of radio galaxies of \cite{VM03},
and for the RLQs of
\citet{H91a, H91b}. The surface brightnesses taken from the literature were all
corrected to a common redshift of $z=4.54$ (the average value of our sample),
so that they may be compared. The FWHM values listed in Table~1 of CJW are local
values, which do not include the velocity field of the gas; in order to
compare with our values, which include a systematic velocity field if any, we
estimated ``global'' FWHM values from the one-dimensional (1D) spectra shown in Fig.~1 of CJW,
which are larger since they include the velocity field. The surface brightnesses
of the objects of \citet{H91a} were estimated from their Table~1, by dividing
the spectral flux $F_\mathrm{s}$ by $1.5\times D_\mathrm{s}$ ($1.5\arcsec$ being
the width of the slit and $D_\mathrm{s}$ the maximum angular extent of the
envelope) when the envelope is detected on one side of the quasar only, and by
dividing $F_\mathrm{s}$ by $3\times D_\mathrm{s}$ when the envelope is detected
on both sides of the quasar.

It seems that the envelopes
of all quasars (whether radio loud or not) and radio galaxies have similar
maximum surface brightnesses. Most of our objects lie below the general trend,
thanks to our much deeper detection limit, and because not all RQQs are
surrounded by bright blobs. On the other hand, the FWHM of the extended \lya\
emission spans almost a factor of ten. Surprisingly, the object in our
sample that has the highest surface brightness, has a wide FWHM, much wider
than those of CJW, and surrounds a RQQ. Our objects with fainter
surface brightnesses have, on average, wider emission lines
than those of CJW. The reason for this is unclear and this result may be a statistical
fluke. Nevertheless, it is interesting to note that the lower left region of
the diagram is observationally more accessible than the lower right region, i.e.
the detection of low surface
brightness envelopes is easier for small FWHMs than for large ones. Therefore,
the relative scarcity of objects in this region cannot be the result of an
observational bias, and must be real. One can only speculate
that the kinematics of these envelopes might be more violent at redshift 4.5 than
later, though such a rapid evolution (recalling that the CJW objects are at
$z\sim 3.2$) seems doubtful.

\subsection{Upper limit to the flux in the N\,\textsc{v} doublet}
The N\,\textsc{v}$\lambda 1238.81-1242.80$~\AA\ doublet is not detected in
emission, in any of the five envelopes detected here in \lya. It may therefore be
interesting to estimate the upper limit to the flux emitted by this doublet,
because this may be translated into an upper limit to the metallicity. We
performed this estimate in the following very simple way. We adjusted the flux
so that the doublet is almost detectable in the 1D spectra of
Fig.~\ref{1Dspectra}, under the assumptions that: 1) the width of the \lya\
line is essentially due to a Maxwellian distribution of the gas velocities, so
that the same width can be adopted for each component of the doublet, 2) the
two components of the doublet have the same strength, and 3) the intensities of
the two components can simply be added. We are aware that our three assumptions are
simplistic, especially the second one, which is equivalent to assuming an
optically thick gas. The weighted oscillator strength $gf$ of the blue
component of the doublet is twice as large as that of the red component
\citep{MW95}, so that the ratio of the blue to red component intensities would
be two in the optically thin case. The latter case would make the detection
easier because the resulting peak would be narrower; conversely, our assumption
implies a wider peak, hence a less easy detection, and so appears rather
conservative.

The results are summarized in Table \ref{Nlimit}, which lists respectively the
upper limit to the flux possibly emitted in the N\,\textsc{v} doublet, the flux
in the \lya\ line, and the ratio of the two. The ratios are not very compelling,
since the lowest is $0.14$, meaning that the N\,\textsc{v} doublet is at least
seven times less strong than the \lya\ line. One sees that, the narrower the
\lya\ line, the less compelling the upper limit on the N\,\textsc{v} doublet,
because the flux of the latter is distributed over a wider wavelength range.

\begin{table}[t!]
\caption{Upper limits to the flux emitted by the envelope in the
N\,\textsc{v} doublet, compared to the flux emitted in the \lya\ line.}
\label{Nlimit}
\centering
\begin{tabular}{lccc}
\hline\hline
Object  & Flux(N\,\textsc{v})& Flux(\lya) &Flux \\
        & \multicolumn{2}{c}{[erg cm$^{-2}$ s$^{-1}$]}& ratio\\
\hline
\br     & $\lesssim2\times 10^{-18}$ & $1.4\times 10^{-17}$ & $< 0.14$\\
\ja     & $\lesssim1\times 10^{-18}$ & $4.3\times 10^{-18}$ & $< 0.23$\\
\jb     & $\lesssim5\times 10^{-17}$ & $3.1\times 10^{-16}$ & $< 0.16$\\
\brb    & $\lesssim1.3\times 10^{-18}$ & $2.6\times 10^{-18}$ & $< 0.50$\\
\hline
\end{tabular}
\tablefoot{ The last
column gives the ratio of column 2 to column 3 (see text).}
\end{table}

\section{Conclusions}

We have performed deep slit spectroscopy of six radio-quiet quasars in the
redshift interval $4.460 \leq z \leq 4.588$ and found extended \lya\ emission
around four of them. The depth of our detection limit is unprecedented, so we
have been able to detect nebulae that are much fainter than in previous surveys.

The main conclusions of this study can be summarized as follows:
\begin{itemize}
\item At $z\approx 4.5$, extended \lya\ envelopes are found around roughly
two-thirds of the quasars.
\item The size of the typical envelope is very large (between $\sim26$~kpc and $\sim64$~kpc),
and does not depend on the \lya\ luminosity of the
BLR of the quasar.
\item The average surface brightness of the envelopes is very low, since it
ranges from $\sim3\times10^{-19}$ to $2\times 10^{-17}$~\esca. Such low values
are due to the large size of the envelopes. Both the large size and the low
surface brightness seem difficult to reconcile with a scenario that attributes
most of the \lya\ luminosity to fluorescence
induced by the quasar \citep{HR01}.
\item The luminosity of the envelopes seems to be correlated with that of
the BLR, but this is not an absolute rule, because of the exceptional behaviour
of \brb. Likewise, the envelope luminosity appears to be roughly correlated with the
predicted ionizing flux of the quasar, but only when \brb\ is excluded. If
confirmed, this trend would resemble that of CJW for RQQs, but over
a wider range of luminosities.
\item The average FWHM of the extended emission line varies from about
$900$~\kms\ to $2500$~\kms\ in our sample alone, while it is often smaller in
other samples (down to $250$~\kms in the radio galaxies of \citealp{VM07}).
\item The N\,\textsc{v}$\lambda 1238.81-1242.80$~\AA\ doublet remains undetected
in all our objects. The most compelling non-detection occurs in \br, where the
flux in the doublet is at least seven times lower than in the \lya\ line.
\end{itemize}
Do these results favour the ``cold accretion'' scenario advocated
by \cite{DL09}, or rather the ``heating'' scenario supported by e.g.
\cite{GALS09}?\footnote{The ``heating'' versus
``cooling'' alternative highlighted by the title of Geach et al.'s paper is
confusing. Indeed, in both models the cold gas is heated, which makes it able to
glow in \lya. Only the source of heating differs. In one case, heating is
provided by the UV radiation of a quasar, while in the other case heating
results from the conversion of gravitational energy into thermal energy; in the
first case, there is photo-ionization, while in the second case there is
collisional excitation.} As discussed above, our results remain ambiguous,
because of the
unexpected behaviour of \brb. Arbitrarily discarding this object, the positive trend between
the ionizing flux and envelope luminosity might be considered as favouring the
heating scenario, but taken at face value our results do not provide
enough evidence for either scenario. A larger sample would enable us to
settle the issue, because a clear correlation between the AGN UV luminosity and
the \lya\ luminosity of the envelope is not expected in the cold flow model. In
any case, future theoretical works will have to take the observed low surface
brightnesses into account. As \cite{HR01} stress in the last sentence of their
paper: ``While a detection of \lya\ fuzz would
provide a direct probe of galaxy formation, nondetections at the level of
$10^{-17}$~\esca would already imply strong constraints." Here, we have provided
two non-detections, but at a deeper level (assuming the envelope does extend over
at least a few arcsec$^2$), and three detections of
envelopes with surface brightnesses significantly lower than $10^{-17}$~\esca.
Only one object was found to have a surface brightness of the order of $10^{-17}$~\esca.

A third possibility might be that the detected emission is not {\sl actual}
emission but rather scattering of broad emission-line photons from the quasar.
This is suggested by the observed relation $L($\lya$)\propto L(BLR)^{1.65\pm0.53}$
(Eq.~\ref{flux_env}), which is compatible to  $L($\lya$)\propto L(BLR)^2$.
The scattered flux is, to zeroth order,
proportional to the BLR luminosity times the amount of matter in the envelope;
if the latter scales with the bulge mass, which is itself proportional to the
supermassive black hole mass, which determines the quasar luminosity, then it
is also proportional
to the BLR luminosity, indeed implying that $L($\lya$)\propto L(BLR)^2$.
Photons are then scattered by cool gas on large scales before reaching the
observer. Despite the velocity shift between the gas and
the quasar redshift (of typically a few hundreds of \kms), there
is enough photon flux from the broad quasar emission line
(whose width is several thousands of \kms) to produce a nebular fuzz.
Admittedly, this argument may prove excessively naive, e.g. because in the
cold flow model, the amount of cold gas increases less rapidly than the halo
mass \citep{KKF09}; the predicted power-law index depends on largely uncertain
properties of quasar environments at $z=4.5$, but a more detailed treatment of
these issues is beyond the scope of this paper.
As for timescales,
the nebular size is $< 1\times 10^6$ light years, hence much shorter than
the typical quasar lifetime, which is thought to be of the order of
$1\times 10^7$ years.
That the brighter the broad emission-line emission from
the QSO, the stronger the nebular emission, is a result that is consistent
with this scattering scenario.
The material responsible for the scattering may originate in tidal tails
and other debris, associated with the presumably interacting
system leading to the onset of quasar activity. This would explain
the asymmetry (of the order of $1-2\arcsec$, corresponding to some ten kpc,
see Fig.~\ref{profile}) and the extent of the emission
\citep[see e.g.][]{MH96, S00}. The observed velocity
difference between the peak emission of the QSO and the
nebula, $\lesssim 1000$~\kms, is rather extreme for a local
merger. Nevertheless, the quasar environments at $z\sim 4.5$ may be more
extreme than those found at lower redshifts.
These quasars may live in proto-cluster like
environments with velocity dispersions of order $1000$~\kms, which resemble
today's richest clusters. The presence of an
additional starburst wind component could contribute to the
signal, as well as the cold gas being accreted onto the host galaxy (i.e. the
cold flow gas).

Additional spectroscopic measurements at the same redshift would be welcome in
order to confirm the correlations found here. However, the crucial data that is badly
needed to make progress in understanding these envelopes  is
narrow-band imaging, which will allow us to obtain much more reliable envelope
shapes, extents, and luminosities.
The latter are crucial discriminating criteria to identify the mechanism
responsible for the envelope emission.

\begin{acknowledgements}
We would like to thank Dr. Lise  Christensen for providing us with the
electronic  form of  the Tables in  CJW  and for useful
discussions.  This study  is supported by  the Swiss National  Science
Foundation (SNSF). We thank the anonymous referee for crucial and
constructive comments, and Dr. Michael Rauch for the clarification of a
technical point.
\end{acknowledgements}

\bibliographystyle{aa}
\bibliography{agn}

\end{document}